\documentclass[superscriptaddress,twocolumn,prb,showpacs,floatfix]{revtex4}
\usepackage[dvips]{graphicx}
\usepackage{epsfig}
\usepackage{psfrag}
\usepackage{amsmath,amsfonts,amssymb,bm,ulem}
\usepackage[usenames]{color}

\begin{document}
\title{Many electron theory of $1/f$-noise in hopping conductivity}
\author{A. L. Burin}
\affiliation{Department of Chemistry, Tulane University, New
Orleans, LA 70118, USA}
\author{B. I. Shklovskii}
\affiliation{William P. Fine Institute of Theoretical Physics,
School of Physics and Astronomy, University of Minnesota,
Minneapolis, MN 55455, USA}
\author{V. I. Kozub}
\affiliation{A. F. Ioffe  Physico-Technical Institute of Russian
Academy of Sciences, 194021 St. Petersburg, Russia}
\affiliation{Argonne National Laboratory, 9700 S. Cass Av.,
Argonne, IL 60439, USA}
\author{Y. M. Galperin}
\affiliation{Department of Physics, University of Oslo, PO Box
1048 Blindern, 0316 Oslo, Norway} \affiliation{A. F. Ioffe
Physico-Technical Institute of Russian Academy of Sciences, 194021
St. Petersburg, Russia} \affiliation{Argonne National Laboratory,
9700 S. Cass Av., Argonne, IL 60439, USA}
\author{V. Vinokur}
\affiliation{Argonne National Laboratory, 9700 S. Cass Av.,
Argonne, IL 60439, USA}
\date{\today}
\begin{abstract}
We show that $1/f$-noise in the variable range hopping regime is
related to transitions of many-electrons clusters (fluctuators)
between two almost degenerate states. Giant fluctuation times
necessary for $1/f$-noise are provided by slow rate of
simultaneous tunneling of many localized electrons and by large
activation barriers for their consecutive rearrangements. The
Hooge constant steeply grows with decreasing temperature because
it is easier to find a slow fluctuator at lower temperatures. Our
conclusions qualitatively agree with the low temperature
observations of $1/f$-noise in p-type silicon and GaAs.
\end{abstract}
\pacs{71.23.Cq, 72.70.+m, 72.20.Ee, 72.80.Sk}
\maketitle

\section{Introduction}

\label{intr}

At low temperatures the variable-range hopping conductivity of
doped semiconductors with strongly localized electrons obeys the
Efros-Shklovskii (ES) law \cite{ShE,ShE1,ShE2}
\begin{equation}
\sigma_{\text{ES}} = \sigma_{0}\exp\left[-\left(\frac{T_{\text{ES}}}{T}\right)^{1/2}\right],
\label{eq:vrh_ShE}
\end{equation}
where the temperature $T_{\text{ES}}$ is defined by the electron-electron
interaction at the localization radius $a$ of electronic states,
\begin{equation}
T_{\text{ES}} = C\frac{e^{2}}{k_{B}\kappa a}.
\label{eq:TES}
\end{equation}
Here $C\simeq 2.7$, $e$ is the electron charge, $k_B$ is the Boltzmann
constant,
and $\kappa$ is the dielectric constant of the semiconductor.
The conductivity behavior Eq. (\ref{eq:vrh_ShE}) is used for example
in ion-implanted silicon (Si:P:B) bolometers working as detectors for
high resolution astronomical X-ray
spectroscopy.\cite{McCammon,McCammon1} An absorption of an X-ray
increases the temperature of the semiconductor and this increase is
detected by the change in its conductivity.
The performance of some bolometers is limited by a $1/f-$ noise, which
obeys the Hooge's law\cite{Hooge,ShMKogan}
\begin{equation}
\frac{\delta \sigma_{\omega}^{2}}{\sigma^{2}}=\frac{\alpha_{H}(\omega,
  T)}{\omega N_{D}},
\label{eq:Hooge}
\end{equation}
where $N_{D}$ is the total number of donors,
\begin{equation}
\delta \sigma_{\omega}^{2} \equiv \int_{-\infty}^{+\infty}\! \! \! dt\,
  e^{i\omega
  t}\langle \delta \sigma(t) \delta \sigma (0)\rangle, \
\delta \sigma(t) \equiv \sigma(t) - \langle \sigma(t)\rangle\, ,
\nonumber 
\end{equation}
and $\langle \cdots \rangle$ denotes ensemble average. The
dimensionless Hooge factor $\alpha_{H}(\omega, T)$ measured for
different doping levels grows by six orders of magnitude with the
decreasing temperature following approximate power law
\cite{McCammon,McCammon1,Weiss}
\begin{equation}
\alpha_{H} \propto T^{-6}\, .
\label{eq:Hooge_exp}
\end{equation}
Ref. \onlinecite{Weiss} gives a strong evidence that the noise is
caused by the electron localization. This work investigates the
bulk Si:P:B semiconductor in the range of dopant concentrations on
both sides of the metal-insulator transition. The increase of the
noise strength by several orders of magnitude, when crossing the
metal-insulator transition from metal-like samples to insulating
samples was observed. The noise intensity continues to increase
with decreasing electron localization radius $a$, what proves the
primal significance of the electron localization in the noise
formation. A similar behavior of the normalized $1/f$-noise power
has been recently reported in the low density hole system of  a
GaAs quantum well.~\cite{GaAs} Thus, these and other
experiments\cite{Lee,Lee2,Savchenko,Savchenko2} show that $1/f-$
noise is one of the manifestations of the complex correlated
electronic state (Coulomb glass) formed by localized electrons
coupled by the long-range Coulomb interaction.

The mechanism of $1/f$-noise in hopping conductivity has been
investigated by several theoretical
groups.\cite{Shkl0,Kozub,Yu1,Shkl1} It was first suggested
\cite{Shkl0} that the $1/f-$noise in the nearest neighbor-hopping
transport is associated with electronic traps, in a way similar to
McWorter's idea of $1/f$-noise in MOSFETs.\cite{McWorter} Each
trap consists of an isolated donor within a spherical pore of the
large radius $r$. Such rare configurations form
\textit{fluctuators}, which have two possible states (empty or
occupied) switching back and forth with the very slow rate defined
by the tunneling rate of electron out or into the pore
\begin{equation}
\nu(r)=\nu_{0}\exp(-2r/a)\, ,
\label{eq:electr_tun}
\end{equation}
where $\nu_{0}\sim 10^{12}$ s$^{-1}$ is the hopping rate
determined by the electron-phonon interaction. When a fluctuator
is occupied the host electron cannot participate in the transport
and thus the transitions of fluctuators change the effective
number of ``charge carriers'' leading to a noise in a
conductivity. The exponential sensitivity of the tunneling rate to
the size $r$ of the trap [Eq. (\ref{eq:electr_tun})] makes the
statistics of trap blinking close to the logarithmically uniform
one leading to $\delta \sigma_{\omega}^{2} \propto 1/\omega$.

The idea of a single-electron fluctuator has been extended to the
regime of the variable range hopping in later works.\cite{Yu1,Shkl1}
These theories require that each fluctuator
characterized by the slow relaxation rate $\nu$ has no neighboring
donors,  which belong to the energy band  around the Fermi level of
the width $E_{\nu} = k_B T\ \ln(\nu_{0}/\nu)$ in the sphere of the radius
$R_{\nu}= (a/2) \ln(\nu_{0}/\nu)$ (here and everywhere below we
choose the chemical potential as a reference point of
energy). According to Ref.~\onlinecite{Shkl1}, the  Hooge parameter is
defined by  the probability of such ``pore'' in four-dimensional
energy-coordinate space, which has the form
\begin{eqnarray}
\alpha_{H}(\omega, T) &\propto& \
\exp\left[-\left(\frac{T}{T_{\text{ES}}}\right)^{3}\left(\ln
    \frac{\nu_{0}}{\omega}\right)^{6}\right]
\nonumber\\
&=&\exp \left[-
  \left(\frac{\ln(\nu_{0}/\omega)}{\ln(\nu_{0}/\nu_{\text{ES}})}
  \right)^{6}\right],
\label{eq:ref_Shklovskii1}
\end{eqnarray}
where $\nu_{\text{ES}}=\nu_{0}e^{-(T_{\text{ES}}/T)^{1/2}}$ is the
rate of typical
hops contributing to Eq. (\ref{eq:vrh_ShE}). At small $\omega$ this
Hooge parameter decreases very fast with decreasing $\omega$. As a
result $1/f-$noise is limited to a relatively narrow frequency
interval $(T_{\text{ES}}/T)^{1/2}<\ln(\nu_{0}/\omega)<(T_{\text{ES}}/T)^{3/5}$,
where dependence of $\alpha_{H}(\omega)$ is still weaker than
$1/\omega$. Observation of $1/f$-noise in a wider range of frequencies
$\omega \ll \nu_{\text{ES}}$ still remains a challenge for the theory.

In this work we suggest the model of many-electron fluctuators, which
possess small relaxation rates $\nu$ with much larger probability then
single-electron traps. A new fluctuator is made of $N$ occupied
(neutral) and $N$ empty (positively charged) donor sites quasi-ordered
into the quasi-cubic lattice with the period $R$. They include only
donors with energies within a band with the width
\begin{equation}\label{UR}
U_R \equiv e^{2}/\kappa R\, ,
\end{equation}
which are required to have somewhat  diminished disorder energies
(see below). We assume that the cluster (Fig.~\ref{fig1_new}) has
two energy minima with energy difference of order of $k_{B}T$
realized by two disordered chess-board configurations (Fig.
\ref{fig1_new}).
Optimizing the ``lattice constant'' $R$ and the
number of electrons $N$ we show below that a $1/f$-noise associated
with such clusters behaves as
\begin{eqnarray}
\alpha_{H}(\omega, T) &\sim&
\exp\left[-\left(\frac{T}{T_{\text{ES}}}\right)^{3/5}\ln^{6/5}
    \frac{\nu_{0}}{\nu} \right]
\nonumber\\
&=&\exp\left[-
  \left(\frac{\ln(\nu_{0}/\omega)}{\ln(\nu_{0}/\nu_{\text{ES}})}
  \right)^{6/5}\right].
\label{eq:ab_ans1}
\end{eqnarray}
At small frequencies the absolute value of the exponent in Eq.
(\ref{eq:ab_ans1}) is apparently much smaller then that for the
single electron trap, Eq. (\ref{eq:ref_Shklovskii1}). As a result,
$1/f-$noise exists in the much broader range of frequencies.
$$(T_{\text{ES}}/T)^{1/2}\ll \ln(\nu_{0}/\omega) \ll
(T_{\text{ES}}/T)^{3}\, .$$ At low temperatures, where
$(T_{\text{ES}}/T)^{1/2}\geq 3$, this range is practically unlimited.

Our approach is similar to the previous analysis of clusters of
many local two-level defects\cite{BK,B1,B2} suggested to interpret
the universal behavior of amorphous solids. Long relaxation times
have been achieved in Ref.~\onlinecite{BK} due to the exponential
increase in the relaxation time with the number of defects
belonging to a single cluster (note that the results of
Ref.~\onlinecite{BK} cannot be directly applied to our system
because they are obtained in the zero temperature limit, i. e.
ignoring thermally activated cluster transitions). The importance
of strong electron-electron correlations and many-electron nature
of fluctuators determining $1/f-$noise was anticipated in
Refs.~\onlinecite{ShMKogan1}.

The manuscript is organized as following. In Sec. \ref{sect:mod1} we
introduce the model of interacting localized electrons in a lightly
doped semiconductor. In Sec. \ref{sect:toymodel} we describe
chess-board clusters and their statistics with respect to energies and
relaxation rates. In Sec. \ref{sect:noirel} we consider the noise in
the variable range hopping conduction, induced by chess-board
clusters. In Sec. \ref{sect:generalize}  we report the generalization
of our results to two-dimensional ($2d$) systems and amorphous
semiconductors, where the variable range hopping conductivity may obey
the Mott law.
\begin{figure}[ptbh]
\begin{center}
\includegraphics[
width=2.4in] {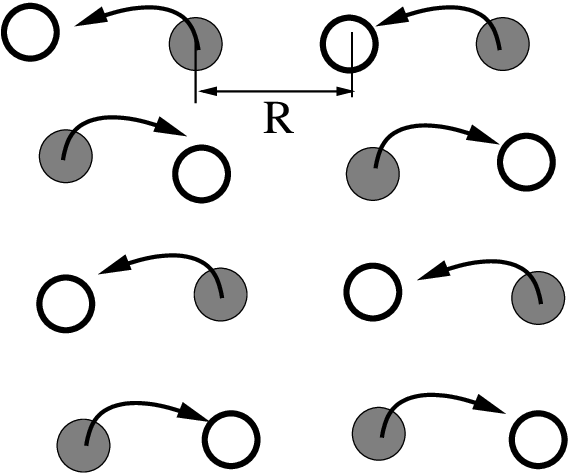} \caption{A ``chess-board'' cluster of
donors, responsible for a $1/f-$
  noise. The dark circles indicate occupied (neutral) donors, while
  the open circles indicate empty donors. All donors are in the energy
  band of the width $2e^{2}/(\kappa R)$ around the chemical potential
  ($E=0$).  Arrows shows the direction of the electron transition
  between two energy minima corresponding to two possible ways to
  occupy cluster sublattices. The length $R$ stands for the size of
  the cluster cell.}
\label{fig1_new}
\end{center}
\end{figure}

\section{Model of a lightly doped semiconductor}
\label{sect:mod1}

Let us consider a $n-$type lightly doped semiconductor with the
concentration $n_{d}$ of donors  and the concentration
$n_{a}<n_{d}$ of acceptors. At low temperatures all acceptors are
charged negatively, $n_{a}$ donors are positively charged, while
$n_{d}-n_{a}$ are occupied and neutral. The set of filling factors
of donors $n_{i}=0, 1$ is determined by the energy minimum of the
Hamiltonian of the classical impurity band\cite{ShE1}
\begin{equation}
\widehat{H}= \sum_{i}
\left[\phi_{i}n_{i}+\frac{1}{2}\frac{e^{2}}{\kappa}\sum_{k\neq
    i}\frac{(1-n_{i})(1-n_{k})}{\mid {\bf r}_{i} - {\bf r}_{k}
    \mid}\right],
\label{eq:Hamiltonian1}
\end{equation}
where $\phi_{i}= \frac{e^{2}}{\kappa}\sum_{j}\frac{1}{\mid {\bf r}_{i}
  - {\bf r}_{j} \mid} $ is the random potential created on the donor $i$ by all acceptors
(enumerated by $j$).
In the ground state energies of donor states are
\begin{equation}
\epsilon_{i} = \phi_{i} - \frac{e^{2}}{\kappa}\sum_{k \neq i}
\frac{1-n_{k}}{\mid {\bf r}_{i}-{\bf r}_{k}\mid}\, .
\label{eq:el_en1}
\end{equation}

It is known \cite{ShE,ShE1} that the density of states $g(\epsilon)$
of such a system has a double peak structure with the soft Coulomb gap
\begin{equation}
g(\epsilon) = \frac{3}{\pi}\frac{\epsilon^{2}\kappa^{3}}{e^{6}}
\label{eq:CoulGapDOS}
\end{equation}
in the middle. At $n_{a} \sim n_{d}/2$ the width of the Coulomb
gap in the density of states is comparable with the total width of
the peak of the density of donor states. It is the Coulomb gap,
which leads to the ES law, Eq. (\ref{eq:TES}). A characteristic
length of hops leading to Eq. (\ref{eq:vrh_ShE}) is
$R_{\text{ES}}=(a/2)(T_{\text{ES}}/T)^{1/2}$ and characteristic
width of the band of energies of states contributing to Eq.
(\ref{eq:vrh_ShE}) is $E_{\text{ES}}=(T_{\text{ES}}T)^{1/2}$ (see
Refs.~\onlinecite{ShE,ShE1,ShE2}).

\section{Long-living chess-board clusters}

\label{sect:toymodel}

\subsection{Chess-Board cluster and its probability}

Chess-board clusters (see Fig. \ref{fig1_new}) are formed by the
rare fluctuations in positions and energies of donors. We require
the following conditions to be satisfied for the cluster. First,
all occupied and empty donors $i$ belonging to the cluster are
placed approximately near the sites $R_{i}$ of a cubic lattice
with the period $R$. This requires
\begin{equation}
\mid {\bf r}_{i} - {\bf R}_{i} \mid < \eta_{R} R\, , \quad \eta_{R}
\leq 1/2\, ,
\label{eq:cluster_coord_restr}
\end{equation}
where $\eta_{R} \sim 1$ is the parameter, restricting the deviations
of electrons and holes belonging to the cluster from the cell
centers.

In addition, all donor sites $i$ of the cluster interact with the
environment. This interaction includes the  disorder energy
$\phi_{i}$ produced by acceptors and the Coulomb interaction with
``external'' empty donors, which do not belong to the cluster,
because they either have energies larger than $U_R$ or spatially
are away from the cluster
\begin{equation}
\varphi_{i}=\phi_{i} + \frac{e^{2}}{\kappa}\sum'_{k}\frac{1-n_{k}}{r_{ik}}
\label{eq:outer_energy1}
\end{equation}
where $\sum'$ indicates that the sum is taken over the ``external''
charged donors only.
For all donors of the cluster disorder energy $\varphi_{i}$ is
required to be smaller than the characteristic energy $U_R$,
Eq~(\ref{UR}),  of the Coulomb interaction within the cluster
\begin{equation}
\varphi_{i} < \eta_{E} U_{R}\, , \quad \eta_{E} < 1\,.
\label{eq:cluster_energy_restr}
\end{equation}
This second requirement involves the energy constraint parameter,
$\eta_{E}$, which should be sufficiently small to keep the two
minimal energy states chess-board like, Fig.~\ref{fig1_new}.

First we evaluate the probability $p_{s}$ for the each chess-board
cell to have the donor, located nearby its center with the
diminished disorder energy. This probability is given by the
product of the density of electronic states within the domain of
the allowed energy, $\int_{0}^{\eta_{E}U_R} g(E)\,dE$, [see
Eq.~(\ref{eq:cluster_energy_restr})] and the allowed volume
$4\pi\eta_{R}^{3}R^{3}/3$,
Eq. (\ref{eq:cluster_coord_restr}). Then using the definition
(\ref{eq:CoulGapDOS}) of the density of states within the Coulomb
gap we get
\begin{equation}
p_{s} = 4\eta_{R}^{3}\eta_{E}^{3}/3\,.
\label{eq:sngl_crstst_strong}
\end{equation}
As discussed previously, $\eta_{E}<1$ and $\eta_{R}<1/2$ in order to
have some ordering of the structure. These two constraints make the
probability $p_{s}$ substantially smaller than 1.
It is convenient to express the probability $p_{s}$ in the exponential
form as $p_{s} = \exp(-\lambda/2)$.
The probability to satisfy constraints
Eqs. (\ref{eq:cluster_coord_restr}), (\ref{eq:cluster_energy_restr})
for all $2N$ sites is given by the product of all $2N$ probabilities
$p_{s}$
\begin{eqnarray}
p_{\text{tot}}=p_{s}^{2N} \approx \exp(-\lambda N)\, .
\label{eq:total_cluster}
\end{eqnarray}
Thus the probability to form the chess-board cluster with two energy
minima separated by the large potential barrier decreases
exponentially with the number of electrons within the cluster.
It is important that within the Coulomb gap the parameter $\lambda$ is
the number of order of unity.

To characterize $1/f$ conductivity noise induced by clusters one has
to define the cluster unit volume probability density, $P(N, \Delta,
R)$,  of a transition energy $\Delta$, a cell size $R$, and a number of
donors $N$. By definition, the product $P(N, \Delta, R)\, dN\,dR\,d\Delta$
characterizes the number of clusters in the unit volume with the cell
size in the domain $(R, R+dR)$, energy in the range $(\Delta,
\Delta+d\Delta)$ and the number of donors in the range $(N, N+dN)$.
The main dependence of the cluster density on the number of donors $N$
is given by the exponent Eq. (\ref{eq:total_cluster}). Next, we should
define the pre-exponential factor. We will see that since the
probability rapidly decreases with increasing $N$ the characteristic
$N$ are not very large and, therefore calculating the pre-exponential
factor we can assume that $N\sim 1$. We ignore the weak
logarithmic dependence of cluster density on the energy $\Delta$ due
to the dipole gap. Then the probability density of clusters, $P(\Delta,
R, N)$, does not depend on $\Delta$ and can be expressed in the way
\begin{equation}
P(N, \Delta, R) \approx \frac{ e^{-\lambda
    N}}{R^{4}U_R}= \frac{\kappa e^{-\lambda N}}{e^{2}R^{3}}\, .
\label{eq:clust_dens1}
\end{equation}
Indeed, the density distribution of the parameters $N$,
$\Delta$, and $R$ is given by the probability of finding a chess-board
cluster having $N$ donors,  $p_{\text{tot}} \propto e^{-\lambda N}$,
divided by the cluster's volume, $NR^3$, and the typical energy
bandwidth of these donors, $U_R\sqrt{N}$. The probability density of
clusters is then
$$P(N, \Delta, R) \propto \frac{\partial}{\partial R}
\frac{p_{\text{tot}}}{N^{3/2} R^3 U_R}\, .$$
Below we will keep track of the cluster size, $N$, only in the
exponent. In this way we arrive at Eq.~(\ref{eq:clust_dens1}).
This expression can be written staring just from the
dimensionality requirement. The function $P(N, \Delta, R)$ cannot
depend on other system parameters because of the universality of
excitations within the Coulomb gap.\cite{ShE,ShE1}

\subsection{Transition rate}

\label{subsect:TA1}

\begin{figure}[t]
\begin{center}
\includegraphics[
width=8cm]
{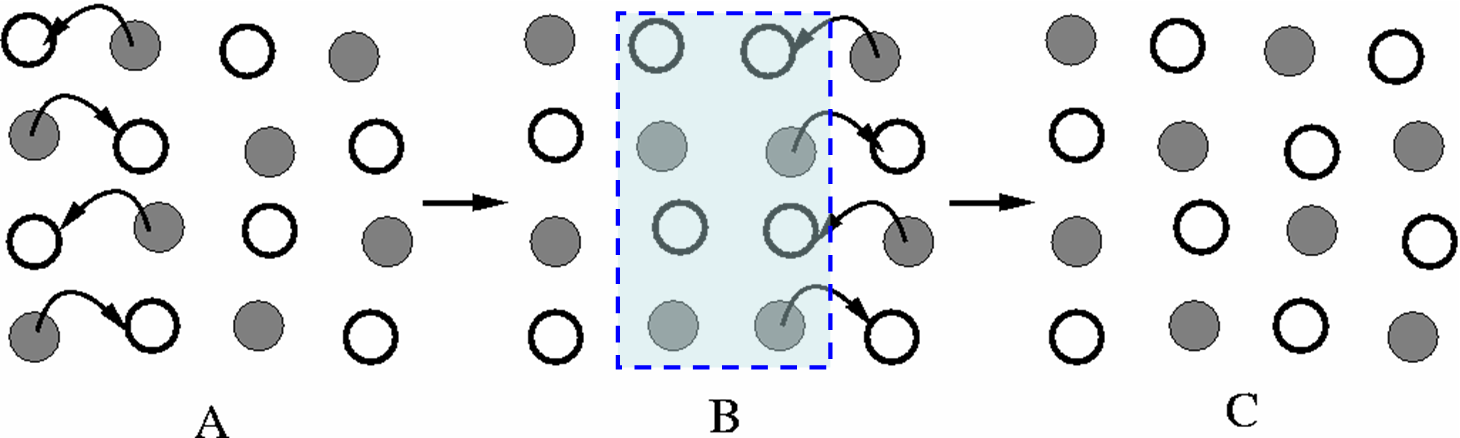}
\end{center}
\caption{Thermally activated transition of the ``chess-board'' cluster
  between two energy minima $A$ and $C$ along the route $A\rightarrow B
  \rightarrow C$.  The activation energy is proportional to the
  number of electrons $N^{2/3}$ in the domain wall shown in the panel
  $B$ by dashed lines. \label{fig5} }
\end{figure}

Transition between two states, $A$ and $C$, of a chess-board
cluster (Fig. \ref{fig5}) can be made via consecutive transition
of single electrons. In the beginning this leads to monotonous
growth of energy until proper saddle point is reached. Thus the
rate of a thermally activated transition between the two states of
fluctuator can be estimated using the minimum activation energy
barrier $E_{A}$ separating two configurations. For the
``chess-board'' structure in Fig.~\ref{fig5} a minimum potential
barrier (saddle point) is defined by the energy of the ``domain
wall'' formation. One has to move all electrons in the one (left
most) column of the cluster $A$ (see Fig.~\ref{fig5}) to the next
positions $B$ to the right side. The activation energy of that
move reads $E_{A} \approx N^{2/3}U_R$, because the number of
electrons belonging to a domain wall is $(N^{1/3})^{2} = N^{2/3}$.
Then this domain wall can move in the system without acquiring or
releasing a significant energy. When the domain wall reaches the
right side the transition is completed. The rate of this process
can be estimated using the standard Arrhenius law with the
activation energy $E_{A}$
\begin{equation}
\nu_{A} = \nu_{0}\, e^{-N^{2/3}U_R/k_{B}T}\, .
\label{eq:genA2}
\end{equation}

An alternative process of transition between energy minima $A$ and
$C$, Fig. \ref{fig5}, is simultaneous tunneling of all electrons
of the cluster. This transition of all $N$ electrons takes place
in the $N$ - th order of the perturbation theory in the weak
tunneling amplitude of an electron $t \propto e^{-R/a}$, so it can
be approximately described by the exponential law
\begin{equation}
\nu_{\text{tun}} \approx \nu_{0}\, e^{-2NR/a}.
\label{eq:gen_tun_1}
\end{equation}
The overall cluster relaxation rate $\nu$ can be approximately
expressed as the sum of the thermally activated and tunneling rates
[Eqs.~(\ref{eq:genA2}), (\ref{eq:gen_tun_1})]:
\begin{equation}
\nu =
\nu_{0}\left[\exp\left(-N^{2/3}\frac{U_R}{k_{B}T}\right)+\exp\left(-2N\frac{R}{a}\right)\right].
\label{eq:tot_rate}
\end{equation}

\subsection{Distribution of clusters over their relaxation rates}
\label{subsect:clusterstat}

The distribution function  of ``chess-board'' clusters over their
relaxation rates $\nu$ (inverse relaxation times) can be
calculated using  Eqs. (\ref{eq:clust_dens1}),
(\ref{eq:tot_rate}). To calculate this distribution function
it is convenient to
  invert the function $\nu(N,R)$ given by Eq. (\ref{eq:tot_rate}) to
  obtain $N(\nu,R)$. The
function $N(\nu,R)$ crosses over from
$N_{\text{ac}}(\nu,R)=\left(k_BT/U_R\right)^{3/2}\ln^{3/2}
  (\nu_0/\nu)$ to $ N_{\text{tun}}(\nu,R)=(a/2R)\ln (\nu_0/\nu)$
at
\begin{eqnarray}
R_{c} &\approx&
a\left(\frac{T_{\text{ES}}}{T}\right)^{3/5}\ln^{-1/5}
\frac{\nu_{0}}{\nu}\, ,
\nonumber\\
N_{c}&\approx& \left(\frac{T_{\text{ES}}}{T}\right)^{-3/5}\ln^{6/5}
\frac{\nu_{0}}{\nu}\, .
\label{eq:R_N1}
\end{eqnarray}
These values correspond to the case when the thermal activation
and tunneling transition rates in Eq. (\ref{eq:tot_rate}) are
equal each other. Indeed, at small $R$ tunneling is more frequent
than  a thermally activated motion. Then according to Eq.
(\ref{eq:tot_rate}) at a given $\nu$ we have $N \propto R^{-1}$.
Since the cluster probability decreases with $N$, it grows with
$R$. This increase takes place until the crossover point $R_{c}$,
where the thermal activation becomes comparable with the
tunneling. Further increase of $R$ will reduce the cluster
probability, because for thermally activated processes the number
of cluster sites grows as $N \sim R^{3/2}$. Therefore, the
quantities $R_{c}$ and $N_{c}$ are the \textit{optimum}
parameters, which  are defined by the crossover between thermal
activation and tunneling regimes
\begin{equation}
\nu/\nu_0 = e^{-2N_{c}R_{c}/a}=e^{-N_{c}^{2/3}e^{2}/\kappa
R_{c}k_{B}T}\, .
\label{eq:cross1}
\end{equation}

The cluster probability density $f(\nu, \Delta)$ can be found using
the previously defined probability density $ P(N, \Delta, R)$,
Eq. (\ref{eq:clust_dens1}), as
\begin{eqnarray}
&&f(\nu, \Delta) = \int_{0}^{\infty} \! \! dR \sum_{N=1}^{\infty}
P(N, R, \Delta) \nonumber\\ && \
\times\delta\left[\nu-\nu_{0}\left(e^{-N^{2/3}U_R/k_{B}T}+e^{-2NR/a}\right)
\right]
\nonumber\\ && \
\approx \frac{1}{\nu R^3_{\text{ES}}
  k_BT_{\text{ES}}}e^{-\lambda N_{c}}
\nonumber\\ && \
= \frac{1}{\nu R^3_{\text{ES}}
  k_BT_{\text{ES}}}\exp\left[-\lambda
    \left(\frac{T}{T_{\text{ES}}}\right)^{3/5}\! \!
    \ln^{6/5}\frac{\nu_{0}}{\nu}\right]. \label{eq:distr4}
\end{eqnarray}
The above equation is valid only if $N_c\gtrsim 1$, which
means $\nu \ll \nu_{\text{ES}}$ and $R < R_{\text{ES}}$.  At $\nu
\sim \nu_{\text{ES}}$ we arrive at $R=R_{\text{ES}}$ and $N_c=1$,
and clusters instead of slow modulating of ES hopping conductivity
become a part of the conducting network of the ES variable range
hopping. As discussed previously, we set $N_c\sim 1$ in the
pre-exponential factor, which yields $R_{c} \approx R_{\text{ES}}$
where $R_{\text{ES}}=a(T_{\text{ES}}/T)^{1/2}$ is ES hopping
length.

One can approximate Eq.~(\ref{eq:distr4}) by the power law
\begin{equation}
f(\nu, \Delta) \propto \frac{1}{\nu}\left(\frac{\nu}{\nu_{0}}\right)^{\beta},
\label{eq:pow_law_weak_int}
\end{equation}
with
\begin{equation}
\beta = -\frac{d\ln f}{d\ln\nu}+1=
\frac{6\lambda}{5}\left(\frac{T}{T_{\text{ES}}}\right)^{3/5}
\ln^{1/5}\frac{\nu_{0}}{\nu} \, .
\label{eq:corr_1f}
\end{equation}

Obviously the statistics $1/\nu$ is applicable if the exponent $\beta
\ll 1$. According to Eq.~(\ref{eq:pow_law_weak_int}) this takes place
at frequencies $\nu \gg \nu_{\min}$, where
\begin{equation}
\nu_{\min} \approx \nu_{0}
\exp\left[-(5/6\lambda)^{5}\left(T_{\text{ES}}/T\right)^{3}\right].
\label{eq:1fnappl}
\end{equation}
In this region the exponential term in
Eq.~(\ref{eq:distr4})depends on $\nu$ only weakly, so the main
dependence is defined by the $1/\nu$-statistics. As we mentioned
in Introduction, $\nu_{\min}$ cannot be practically distinguished
from zero at low temperature.

Note that the realistic variable range hopping measurements are
done typically close to metal-insulator transition, where
$n_{d}a^{3} \simeq 1$. In this case, the inequality $\nu \gg
\nu_{min}$ always means that $R \gg n_{d}^{-1/3}$ and $U_R$ does
not exceed the width of the Coulomb gap. However, if in the
lightly doped limit $n_{d}a^{3} \ll 1$ strictly speaking the
distribution Eq. (\ref{eq:distr4}) is applicable within the
Coulomb gap only, when $R>n_{d}^{-1/3}$. It can be shown that the
case $a<R<n_{d}^{-1/3}$, where $U_R$ exceeds the width of the
Coulomb gap, only a minor revision of our theory is needed.
Namely, the parameter $\lambda$ should be increased by a
logarithmically large factor.

\section{Conductivity noise}
\label{sect:noirel}

Below we discuss how cluster transitions induce a $1/f$-noise of
the electric current. The hopping conductivity
Eq.~(\ref{eq:vrh_ShE}) is provided by the critical network of
resistances percolation cluster.\cite{ShE1} According to the
Shklovskii-De Gennes model of infinite percolation
cluster\cite{ShE1} one can view this network as the random lattice
formed by one-dimensional links of the approximate length $R_{T}
\sim a T_{\text{ES}}/T$. Each link is a chain of
$\sqrt{T_{\text{ES}}/T}$ donor sites separated by the hopping
length $R_{\text{ES}} \sim a \sqrt{T_{\text{ES}}/T}$.

Following the ideas of Kozub,\cite{Kozub} we consider the conduction
noise induced by the cluster transitions affecting adjacent
links.
A link has the critical hop with the largest resistance $r_{h}
\propto e^{\sqrt{T_{\text{ES}}/T}}$ which is comparable with the
resistance of the whole chain. The noise is defined by the
clusters located in the vicinity of these critical sites. Assume
that the critical hop  is separated from the nearest neighbor
cluster by the distance $R$ and this cluster makes slow
transitions between its two energy minima. The transition of the
cluster changes its characteristic dipole moment by the value $\mu
\sim eRN^{1/2} \sim e R_{\text{ES}}$. The chess-board dipole
potential leads to the energy fluctuation of the critical hop
$\delta E \sim e^{2}R_{\text{ES}}/R^{2}$. If this energy exceeds
the thermal energy $T$, or
\begin{equation}
R < R_{\text{int}} = a\left(T_{\text{ES}}/T\right)^{3/4}\, ,
\label{eq:dist_to_cluster}
\end{equation}
the resistance of the critical hop changes substantially, leading to
the addition or removal of the whole link. The fluctuation of the
sample conductivity induced by a change of one link can be expressed
as
\begin{equation}
\delta \sigma \sim \frac{\sigma_{\text{ES}}}{N_{l}} =
\sigma_{\text{ES}} \frac{R_{T}^{3}}{V},
\label{eq:noise11}
\end{equation}
where $V$ is the system volume and the ratio $N_{l} \approx
V/R_{T}^{3}$ estimates the total number of links.

Only clusters with the energy $\Delta$ comparable to the thermal
energy are able to contribute to the noise, while the contribution
of others is exponentially suppressed. Therefore, the relevant
clusters density for the noise at  frequency $\omega$ is
$n_{\omega}\approx k_B T\, f(\omega, k_BT )$, where $f(\omega,
k_BT)$ is given by Eq. (\ref{eq:distr4}). The probability $p_{l}$
that the given link has the cluster located nearby, Eq.
(\ref{eq:dist_to_cluster}), can be expressed through the cluster
density as $p_{l} \approx f(\omega, k_BT)\, k_BT\,
R_{\text{int}}^{3}$. Then the noise at the frequency $\omega$ is
induced by $N_{\omega}\sim p_{l}N_{l}$ clusters making random
contributions. Using Eq. (\ref{eq:noise11}) one can express
the noise intensity as
\begin{equation}
\frac{\delta \sigma_{\omega}^{2}}{\sigma_{\text{ES}}^{2}}
\sim\frac{N_\omega}{N_l^2} =  \frac{p_{l}}{N_{l}}\, .
\label{eq:noise1}
\end{equation}

Using the latter equation one arrives at the Hooge parameter
\begin{eqnarray}
\alpha_{H}=n_{d}a^{3}\left(\frac{T_{\text{ES}}}{T}\right)^{11/4}\! \!\!
\exp\left[-\lambda \left(\frac{T}{T_{\text{ES}}}\right)^{3/5}\!\!\!
\ln^{6/5}\frac{\nu_{0}}{\nu}\right] .
\label{eq:Hooge2ans}
\end{eqnarray}
 The main exponential
term of Eq. (\ref{eq:Hooge2ans}) apparently agrees with Eq.
(\ref{eq:ab_ans1}). According to this equation, at low temperature
the prefactor of $\alpha_{H}$ is much larger than unity. This
happens because standard normalization of $\alpha_{H}$ to the
total number of donors (or electrons) is not natural for the
variable-range hopping, where only small fraction of all donors
participates in transport.

It is the straightforward consequence of the cluster distribution
given by Eqs. (\ref{eq:distr4}), (\ref{eq:Hooge2ans}) that for ES
hopping, the fluctuators, defined by the single electron
pores\cite{Shkl1} have much smaller density (see Eq.
(\ref{eq:ref_Shklovskii1})) and they can be always neglected. Note
that this is not the case for  Mott variable-range hopping as
discussed in Sec. \ref{sect:generalize}.

In the derivations of Eqs. (\ref{eq:distr4}) and
(\ref{eq:Hooge2ans}) we assumed that
$E_{\text{ES}}=k_B\sqrt{TT_{\text{ES}}}\ll
e^{2}n_{d}^{1/3}/\kappa$,
$R_{\text{ES}}=(a/2)(T_{\text{ES}}/T)^{1/2}\gg n_{d}^{-1/3}$, and
the variable range hopping takes place. If these conditions are
violated and $k_BT\gg e^{2}n_{d}^{2/3}a/\kappa$ one deals with the
nearest neighbor hopping conductivity, where the Coulomb
interaction plays a secondary role, see  Ref.~\onlinecite{ShE1}.
In this case there is a range of not very small frequencies
$$\frac{1}{n_{d}^{1/3}a}<\ln\frac{\nu_{0}}{\omega}<\frac{1}{n_{d}^{1/3}a}\left(\frac{\kappa
    k_{B}T}{e^{2}n_{d}^{2/3}a}\right)^{1/3},$$ where
$1/f$-noise is determined by single-electron pores.\cite{Shkl0} At
very small frequencies again the chess-board clusters take over.

Returning to the variable range hopping we can generalize our
theory to a two-dimensional system. In this case the density of
states is proportional\cite{ShE2} to $|\epsilon|$ rather than to
$\epsilon^2$. In addition, the ``domain wall'' energy is
proportional to $\sqrt{N}$, while the cluster volume is
proportional to $R^2$. Taking these canges into account and
following the previous derivation we arrive at the expression
\begin{eqnarray}
\alpha_{H}^{(2d)}(\omega, T) \propto \exp\left[-\lambda
  \left(\frac{T}{T_{\text{ES}}}\right)^{2/3}\ln^{4/3}\frac{\nu_{0}}
{\omega}\right].
\label{eq:Hooge2dans}
\end{eqnarray}
We see that in $2d$ the $1/f$-noise amplitude $\alpha_{H}$ is
smaller than in $3d$. This happens because of smaller activation
energy in $2d$ systems compared to $3d$ systems (see Sec.
\ref{subsect:TA1}). Similarly to a $3d$ case, in $2d$ a deviation
from $1/f$-behavior becomes large at very  small frequencies. The
criterion for $1/f$-noise reads
\begin{equation}
\nu_{\text{ES}}\gg \omega\gg \omega_{\min} \approx \nu_{0}
\exp\left[-\left(\frac{3}{4\lambda}\right)^{3}\left(\frac{T_{\text{ES}}}{T}\right)^{2}\right]\,
.
\label{eq:1fnappl2d}
\end{equation}
Although the low-frequency cutoff $\omega_{\min}$ exceeds that for a
$3d$ system  it is still hardly distinguishable from zero at
temperatures $T\lesssim T_{\text{ES}}/5$.

Our theory can not lead to an observable $1/f$-noise in 1d case,
because in this case the domain wall contains one electron, its energy is small and it is practically 
impossible to construct a
slow enough trap at a finite temperature.

\section{Mott variable range hopping.}

\label{sect:generalize}

In this section we briefly discuss the  case of an amorphous
semiconductor, where the bare density of states is diminished by a
strong compositional disorder of non-Coulomb nature, so that the
Coulomb gap is relatively narrow and the variable range hopping
conductivity obeys the Mott law $\ln \sigma_{\text{M}} \sim
(T_{\text{M}}/T)^{-1/(d+1)}$\cite{Mott}.

For this purpose we depart from the original model Eq.
(\ref{eq:Hamiltonian2}) and consider the  Efros lattice
model\cite{Efros3} characterized by Hamiltonian with a strong
non-Coulomb disorder
 \begin{equation}
\widehat{H} = \sum_{i} \phi_{i}n_{i} + \frac{1}{2} \sum_{ij}
\frac{e^{2}}{\kappa r_{ij}}n_{i}n_{j}\,.
\label{eq:Hamiltonian2}
\end{equation}
Here  $n_{i}=\pm 1/2$ stand
for a hole and an electron, respectively. The
density of bare states $\phi_{i}$ is defined as
$g_{0}=n/2W$, where $n$ is the concentration of lattice sites $i$
and $W$ is the characteristic energy of disorder, i. e.,  random
potentials $\phi_{i}$ are distributed uniformly within the domain
$(-W, W)$. We assume that the dimensionless ratio
$A=W\kappa/e^{2}n^{1/d}$ is much larger than unity. In this case the
Coulomb gap is much narrower than the width of the density of states
$W$.

The results of our theory are summarized in Table~\ref{tab:1}. The
left half of this table characterizes the variable-range hopping
at different temperatures and dimensionalities $d=2$, and $3$ for
both the classical impurity band model and the Efros model. At low
temperatures the variable range hopping conductance obeys the ES
law, Eq. (\ref{eq:vrh_ShE}), while at higher temperature it obeys
the Mott law.

The right half of Table I compares contributions from chess-board
clusters, $\alpha_{H}$, and one-electron pores,
$\alpha_{H}^{\text{pore}}$. As we mentioned above the pores are
not important within the range of validity of the ES law. However,
they become competitive in the case of the Mott law. Remarkably,
in spite of the fact the Coulomb interaction is irrelevant to the
Mott variable range conductivity, the very low frequency
$1/f$-noise is still determined by the chess-board clusters bound
by the Coulomb interaction. Naturally single electron pores
dominate at higher frequencies close to the frequency of Mott
hops. They also become more important with decreasing
dimensionality because in a low-dimensional system it is easier to
create a pore, but it is more difficult to create a slow
chess-board cluster. Therefore, the range applicability of pores
is broader in the $2d$ case.

\section{Comparison of the theory with the experiment}
\label{exp}

Our analysis predicts the strong temperature increase of the
$1/f$-noise at low temperatures due to the exponential factor
$e^{-F(\omega, T)}$. This expectation qualitatively agrees with
the experiments\cite{McCammon,McCammon1,Weiss}, where the Hooge
parameter is approximately proportional to $T^{-6}$.

As an example, let us discuss the results of experiments on ion
implanted Si:P:B\cite{McCammon,McCammon1} showing the ES
temperature dependence of the conductance. In this material  the
Hooge  parameter, $\alpha_{H}$, has been measured as a function of
temperature in the wide range from $1.4$ K to $44$ K for different
levels of doping leading to different temperatures
$T_{\text{ES}}$. According to Ref.~\onlinecite{McCammon}, behavior
of the Hooge parameter can be described by the empirical law
\begin{equation}
\alpha_{H}=0.034 \,
T_{\text{ES}}^{2.453}\left(T/0.153\right)^{-5.2-0.9
\log_{10}T_{\text{ES}}}\, .
\label{eQ:empiriclaw}
\end{equation}
Here $T$ and $T_{\text{ES}}$ are measured in Kelvin degrees. In
particular, in the material  with $T_{\text{ES}}\approx  11$ K
the relative resistance fluctuations at  $\omega \sim 1$
Hz increased from $10^{-11}$ to $10^{-7}$ as temperature decreased
from 0.3 to 0.008 K.

\begin{figure}[b]
\begin{center}
\includegraphics[
width=3in]
{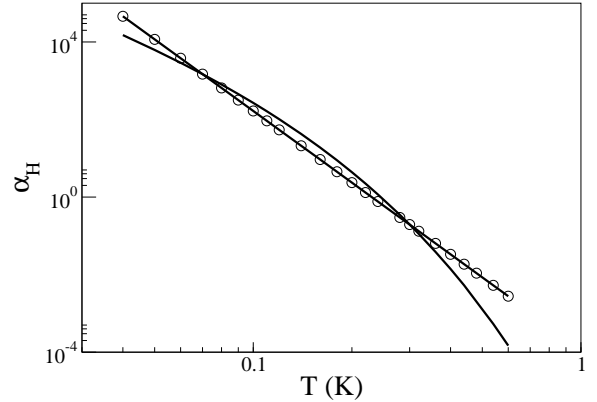}
\end{center}
\caption{Comparison of the theory (solid line,
  Eq.~(\protect{\ref{eq:Hooge2ans}}) and the experiment (solid line with
  circles, Eq. (\protect{\ref{eQ:empiriclaw}}) for the case
  $\omega/2\pi =1$ Hz and
  $T_{\text{ES}}=11$ K. We
  have used $\lambda=2.395$ and $n_{d}a^{3}=0.0084$ to make the
  optimum data fit in the range of temperatures (0.07-0.3 K)
  studied in the experiment. \label{fig7}}
\end{figure}

Let us compare this empirical law with Eq.~(\ref{eq:Hooge2ans})
derived in the previous section. Note that while deriving that
equation we have neglected the powers of $N_c$ in the
pre-exponential factors since $N_c$ cannot be much larger than
unity because of the factor $e^{-\lambda N_c}$, see
Eq.~(\ref{eq:distr4}). It is still important to take into account
the temperature dependence of the minimum transition rate
$\nu_{0}$ because the cluster transitions are induced by their
interaction with phonons. We assume that $\nu_{0}\propto T^{3}$
and take it in the form $\nu_{0}=10^{12}T^{3}$, where $T$ is
measured in Kelvin degrees,  to have $\nu(T) \approx T/\hbar$ at
the energy comparable to the Debye energy. Then for the sample
with $T_{\text{ES}}=11$ K one can obtain a quite reasonable fit of
the experimental data setting $\lambda=2.395$ and
$n_{d}a^{3}=8.4\times 10^{-3}$, as shown in Fig.~\ref{fig7}.
\bigskip
The estimate for the dimensionless constant $\lambda$ agrees with
the expectation $\lambda/2 \sim 1$. The value of $n_{d}a^{3}$
seems to be too small for the material used in the
experiments\cite{McCammon,McCammon1} performed in the vicinity of
the metal-insulator transition. However, our fit is not
sufficiently accurate in interpretation of pre-exponential
factors. In particular, the fitted factor $n_{d}a^{3}$ can be
significantly underestimated due to neglecting of the
$N_c$-dependent pre-exponential factor. What can be even much more
important for the deeper understanding of $1/f$ noise in noisy
samples is the fact that the density of dopants in them is not
uniform~\cite{McCammon1}, but depends on the distance from the
surface.

\begin{widetext}
\begin{center}
\begin{table}[h]
\begin{tabular}
{|c|c|c|c|c|c|c|c|}\hline
$d$ & $T$ & $g(\epsilon)$ & $-\ln \sigma/\sigma_{0}$ & $T_{\text{M/ES}}$ &
$-(1/\lambda)\ln \alpha_{H}$ & $-\ln \alpha_{H}^{\text{pore}}$ & $\alpha_{H}
< \alpha_{H}^{\text{pore}}$
\\ \hline
$2$ & $T<\frac{T_{\text{ES}}^{3}}{T_{\text{M}}^{2}}$ &
$\frac{2}{\pi}\frac{\epsilon \kappa^{2}}{e^{4}}$ &
$\left(\frac{T_{\text{ES}}}{T}\right)^{1/2}$ &
$T_{\text{ES}}\sim\frac{e^{2}}{k_B\kappa a}$ &
$\left(\frac{T}{T_{\text{ES}}}\right)^{2/3}\ln^{4/3}\frac{\nu_{0}}{\omega}$
&
$\left(\frac{T}{T_{\text{ES}}}\right)^{2}\ln^4 \frac{\nu_{0}}{\omega}$ 
& n/a 
\\ \hline
$2$ & $\frac{T_{\text{ES}}^{3}}{T_{\text{M}}^{2}}<T$ & $g_{0}$ &
$\left(\frac{T_{\text{M}}}{T}\right)^{1/3}$ &
$T_{\text{M}}\sim\frac{1}{k_Bg_{0}a^{2}}$ & $
\left(\frac{T}{T_{\text{ES}}}\right)^{2/3}\ln^{4/3}\frac{\nu_{0}}{\omega}$
& $\frac{T}{T_{\text{M}}}\ln^3 \frac{\nu_{0}}{\omega}$ &
$\ln\frac{\nu_{0}}{\omega}<\left(\frac{T_{\text{M}}^{3}}{TT_{\text{ES}}^{2}}
\right)^{1/5}$  
\\ \hline
$3$ & $T<\frac{T_{\text{ES}}^{2}}{T_{\text{M}}}$ &
$\frac{3}{\pi}\frac{\epsilon^{2}\kappa^{2}}{e^{4}}$ &
$\left(\frac{T_{\text{ES}}}{T}\right)^{1/2}$ &
$T_{\text{ES}}\sim\frac{e^{2}}{k_B\kappa a}$ &$
\left(\frac{T}{T_{\text{ES}}}\right)^{3/5}\ln^{6/5}\frac{\nu_{0}}{\omega}$ 
&
$\left(\frac{T}{T_{\text{ES}}}\right)^3 \ln^6 \frac{\nu_{0}}{\omega}$
& n/a
\\ \hline
$3$ & $\frac{T_{\text{ES}}^{2}}{T_{\text{M}}} < T$ & $g_{0}$ &
$\left(\frac{T_{\text{M}}}{T}\right)^{1/4}$ &
$T_{\text{M}}\sim\frac{1}{k_B g_{0}a^{3}}$ & $
\left(\frac{T}{T_{\text{ES}}}\right)^{3/5}\ln^{6/5}\frac{\nu_{0}}{\omega}$  
& $\frac{T}{T_{\text{M}}}\ln^4 \frac{\nu_{0}}{\omega}$
&
$\ln\frac{\nu_{0}}{\omega}<\left(\frac{T_{\text{M}}^{5/2}}{TT_{\text{ES}}^{3/2}}\right)^{1/7}$ 
\\ \hline
\end{tabular}
\caption{
Electronic density of states, parameters of the variable range hopping
in $2d$ and $3d$ systems, $1/f$-noise produced by chess-board clusters
and single pore electrons and their comparison.\label{tab:1}}
\end{table} 
\end{center}
\end{widetext}

\section{Conclusions}
\label{sect:concl}

In this manuscript we have suggested a novel mechanism of the
$1/f$-noise in doped semiconductors in the hopping regime. This
mechanism is associated with many-electron transitions of the
chess-board clusters, the rate of which decreases exponentially with
the cluster size. This exponential dependence results in the close to
$1/\nu$ distribution of clusters over their relaxation rates,
$\nu$. The slow fluctuations of cluster states modulate the
critical resistors forming the backbone cluster  leading to the
$1/f-$noise in the electronic conductivity.

Our predictions regarding the magnitude and temperature dependence of
the Hooge parameter, $\alpha_{H}$, are at least in a qualitative
agreement with experiments in ion-implanted silicon
(Si:P:B).\cite{McCammon,McCammon1} In particular, both specific
features -- dramatic decrease of the noise with
the temperature increase and its increase with the increase of
$T_{\text{ES}}$ -- are explained. Moreover, the experimental results
can be fitted quantitatively using reasonable values of the adjustable
parameters.

The low-temperature ($T < 0.2$ K) experimental values of the Hooge
parameter exceed unity. This fact also agrees with Eq.
(\ref{eq:Hooge2ans}), since at low temperatures the main
temperature dependence is given by the prefactor. Experimental
results obtained in Refs.~\onlinecite{Weiss,GaAs} are also in a
qualitative agreement with our theory.

We are grateful to A. L. Efros, D. McCammon, M. Moore, M. Mueller,
B. Spivak and C. C. Yu for useful discussions.
AB greatly appreciates the hospitality of William P. Fine Theoretical
Physics Institute of
the University of Minnesota after the hurricane disaster in the city
of New Orleans. The work of AB is supported by the Louisiana Board of
Regents (Contract No. LEQSF (2005-08)-RD-A-29) and NSFEpsCor LINK
Program through the Louisiana Board of Regents. The work of YG, VK and
VV was supported by the U. S. Department of Energy Office of Science
through contract No. W-31-109-ENG-38.

\end{document}